\documentclass[preprint,prd,aps,showpacs,showkeys,nofootinbib]{revtex4-1}
\usepackage{amssymb}
\usepackage{amsmath}
\usepackage{graphicx}
\usepackage{subfigure}
\usepackage{float}
\usepackage{mathrsfs}

\usepackage{amsmath}
\usepackage{epstopdf}
\usepackage{dcolumn}
\usepackage{bm}
\usepackage{subfig}

\newcommand{\be}{\begin{eqnarray}}
\newcommand{\ee}{\end{eqnarray}}
\newcommand{\PP}{{\cal P}_{\cal R}}
\UseRawInputEncoding
\begin{document}


\title{Double peaks of gravitational wave spectrum induced from inflection point inflation}

\author{Tie-Jun Gao$^1$}
\email{tjgao@xidian.edu.cn}
\author{Xiu-Yi Yang$^{2}$}
\affiliation{$^1$School of Physics and Optoelectronic Engineering, Xidian University, Xi'an 710071, China\\
$^2$School of science, University of Science and Technology Liaoning, Anshan 114051, China}

\begin{abstract}
We investigate the possibility to induce double peaks of gravitational wave(GW) spectrum from primordial scalar perturbations in inflationary models with three inflection points.
Where the inflection points can be generated from a polynomial potential or generated from Higgs like $\phi^4$ potential with the running of quartic coupling. 
In such models, the inflection point at large scales predicts the scalar spectral index and tensor-to-scalar ratio consistent with current CMB constraints, and the other two inflection points generate two large peaks in the scalar power spectrum at small scales, which can induce GWs with double peaks energy spectrum.
 We find that for some choices parameters the double peaks spectrum can be detected by future GW detectors, and one of the peaks around $f\simeq10^{-9}\sim10^{-8}$Hz can also explain the recent NANOGrav signal.
Moreover, the peaks of power spectrum allow for the generation of primordial black holes, which account for a significant fraction of dark matter.
\end{abstract}

\maketitle

\section{Introduction \label{sec:intr}}

Gravitational wave(GW) astronomy is began since the detection of GWs from mergers of black holes or neutron stars by the LIGO and Virgo collaborations~\cite{ref1,ref2,ref3,ref4,ref5,ref6}.
Besides these from mergers, the GWs can also be generated from perturbation theory of inflation. However, the current constraint of the tensor-to-scalar ratio on CMB scales is $r<0.064$ at $95\%$ confidence level~\cite{ref7} by the Planck 2018 data in combination with BICEP2/Keck Array, which is too small to be detected in the near future.

Although at first order in perturbation theory the scalar and tensor perturbations are decoupled, however, at second order they are coupled, so the second-order GWs can be induced from scalar perturbations when it renter the Hubble radius in the radiation-dominated(RD) era~\cite{ref8,ref9}. In most of the inflationary models, the induced second-order GWs is generically negligible compared to the first-order GWs. However, if the power spectrum of scalar perturbations is enhanced at small scales, the second-order GWs can be sizable or even larger than the first-order one~\cite{ref10,ref11,ref12,ref13,ref14,ref15,ref16,ref17,ref171,ref172,ref173,ref174,ref175}.

One way to realize the enhancement of the power spectrum at small scales in single field inflation is using an inflection point~\cite{ref18,ref19,ref20,ref21,ref22}. The inflection point is such point that both the first and second order derivatives of the potential vanish. Near the inflection point the Hubble slow-roll parameter $|\eta_H|>3$, so the slow-roll approximation fails and the ultra-slow-roll trajectory supersedes, which  gives rise to a large peak in the scalar power spectrum at small scales, and induces GWs with a peak in the energy spectrum. Such inflection point can be generated in critical Higgs inflation with the running of a large non-minimal coupling~\cite{ref23,ref231,ref24}, or generated in the framework of supergravity~\cite{ref25,ref26} or string theory~\cite{ref27,ref28,ref29,ref30} etc.

The previous models with single inflection point only lead to one peak in the scalar power spectrum, with the width about $20$ e-folding numbers. As we all know that the e-folding numbers during inflation is about $50-60$, so it seems possible to generate two peaks in the power spectrum using inflection points. Thus in this paper, we shall investigate the possibility to induce GW spectrum with double peaks by the scalar power spectrum which have two peaks at small scales. We show that such kinds of power spectrum can be generated from the potential with three inflection points, which is realized in a model with polynomial potential or in Higgs like $\phi^4$ potential with the running of quartic coupling by radiative corrections. By fine-tuning the parameters of the models, we show that the double peaks GW signal can be detected by SKA, LISA and other detectors in near future. Recently, the North American Nanohertz Observatory for Gravitational Waves(NANOGrav) report the hint of the stochastic GW signal in its 12.5-year observation of pulsar timing array(PTA), which can be  
fitted by a power-law $\Omega_{GW}\propto f^{5-\gamma}$
with the exponent $5-\gamma\in (-1.5, 0.5)$ at $1\sigma$ confidence level\cite{Arzoumanian:2020vkk,Vaskonen:2020lbd,Kohri:2020qqd,DeLuca:2020agl,Kawasaki:2021ycf}.
We found that one of the peaks of power spectrum at $k\sim10^{6}\mathrm{Mpc^{-1}}$ in our models can explain the NANOGrav signal.

Moreover, the enhancement of scalar perturbations at small scales leads to the production of primordial black holes(PBHs) via gravitational collapse in the
radiation-dominated era\cite{Yokoyama:1995ex,GarciaBellido:1996qt,Clesse:2015wea,Garcia-Bellido:2016dkw,Cheng:2016qzb,Fu:2019ttf}. Although various observations have constrained the fraction of dark matter(DM) in the form of PBHs, there still exist open windows where PBHs are possible to present a significant fraction or even all the DM in our Universe\cite{ref301,ref302,ref303,ref304,ref305,ref306,ref307,ref308,ref309,ref3010,ref3011,ref3012,ref3013,ref3014,ref3015,ref3016,ref3017,ref3018,ref3019,ref3020,ref3021,ref3022}. So we also estimate the generation of PBHs, and get the mass distribution  and abundance of PBHs, which can account for a dominant component of dark matter.

The paper is organized as follows.
In the next section, we brief review the mechanism of the GWs induced by first-order scalar perturbations.
 Then in Sec.3, we setup two inflationary models with three inflection points which can induce GWs with double peaks spectrum.
 The numerically results of the inflaton dynamics, the energy spectrum of induced GWs, and the mass distribution and abundance of PBHs in the two models are presented in Sec.4.
The last section is devoted to summary.


\section{Gravitational waves induced by scalar perturbations  \label{sec:gw}}
In this section, we shall brief review the computation of induced GWs, for more details in Ref.\cite{ref31,ref32,ref33,ref34,ref35,ref36}. In the conformal Newtonian gauge, the perturbed metric is written as
\begin{equation}
\mathrm{d} s^{2}=-a^{2}(1+2 \Psi) \mathrm{d} \eta^{2}+a^{2}\left[(1-2 \Psi) \delta_{i j}+\frac{1}{2} h_{i j}\right] \mathrm{d} x^{i} \mathrm{d} x^{j},
\end{equation}
where $\Psi $ is the scalar perturbations, and the Fourier components of tensor perturbations $h_{ij}$  are defined as usual by
\begin{equation}
\label{eq:fourier}
h _ { i j } (\eta, \mathbf{x}) = \int \frac { d^3 \mathbf{k} } { ( 2 \pi ) ^ { 3 / 2 } } e ^ { i \mathbf{k} \cdot \mathbf{x} } \left[ h_\mathbf{k}^ { + } ( \eta ) \mathrm{e}_{ij}^+ ( \mathbf{k} ) + h_\mathbf{k}^\times ( \eta ) \mathrm{e}_{ij}^\times (\mathbf{k}) \right],
\end{equation}
where the two polarization tensors are
\begin{equation}
\begin{array}{l}{e_{i j}^{(+)}(\mathbf{k})=\frac{1}{\sqrt{2}}\left[e_{i}(\mathbf{k}) e_{j}(\mathbf{k})-\overline{e}_{i}(\mathbf{k}) \overline{e}_{j}(\mathbf{k})\right]}, \\
 {e_{i j}^{( \times)}(\mathbf{k})=\frac{1}{\sqrt{2}}\left[e_{i}(\mathbf{k}) \overline{e}_{j}(\mathbf{k})+\overline{e}_{i}(\mathbf{k}) e_{j}(\mathbf{k})\right]},
 \end{array}
\end{equation}
with the basis vectors $e_{i}(\mathbf{k})$ and $\overline{e}_{i}(\mathbf{k})$ orthogonal to each other and to $\mathbf{k}$. In the following, we shall omit the polarization index for simplicity.

In the Fourier space, the equation of motion of tensor modes can be derived from the Einstein equation as
\begin{equation}
h_{\mathbf{k}}^{\prime \prime}(\eta)+2 \mathcal{H} h_{\mathbf{k}}^{\prime}(\eta)+k^{2} h_{\mathbf{k}}(\eta)=S_{\mathbf{k}}(\eta),
\end{equation}
where $S_{\mathbf{k}}(\eta)$ is the Fourier transformation of the source term,
\begin{equation}
S_{\mathbf{k}}(\eta)=4 \int \frac{d^{3} p}{(2 \pi)^{3 / 2}} \mathrm{e}_{i j}(\mathbf{k}) p_i p_j \Big(2 \Psi_{\mathbf{p}} \Psi_{\mathbf{k-p}}+\frac{4}{3(1+w) \mathcal{H}^{2}} \left(\Psi^{\prime}_{\mathbf{p}}+\mathcal{H} \Psi_{\mathbf{p}}\right) \left(\Psi^{\prime}_{\mathbf{k-p}}+\mathcal{H} \Psi_{\mathbf{k-p}}\right)\Big).
\end{equation}
The scalar perturbations $\Psi_{\mathbf{k}}$ can be split into the primordial value $\psi_{\mathbf{k}}$ and the transfer function $\Psi(k\eta)$
\begin{equation}
\label{eq:split}
\Psi_{\mathbf{k}} \equiv\Psi(k \eta) \psi_{\mathbf{k}}.
\end{equation}
If the peak mode enters the horizon in the RD era, the equation of state is $\omega=1/3$,
and then the transfer function becomes
\begin{equation}
\label{eq:solutontos}
\Psi(x)=\frac{9}{x^{2}}\left(\frac{\sin (x / \sqrt{3})}{x / \sqrt{3}}-\cos (x / \sqrt{3})\right).
\end{equation}

For the modes well inside the horizon, the density parameter of the GWs within the logarithmic interval of the wave numbers can be expressed in terms of the power spectrum of GWs
\begin{equation}
\label{eq:OmegaGW}
\Omega_{\mathrm{GW}}(\eta, k)\equiv\dfrac{1}{\rho_{c}}\dfrac{d\rho_{\mathrm{GW}}}{d\ln k} = \frac{1}{24}\left(\frac{k}{\mathcal{H}}\right)^{2} \overline{{\mathcal P}_{h}(\eta, k)}\,,
\end{equation}
with $\rho_c$ is the critical energy density of the University, the overline denotes the oscillation average and the two polarization modes are summed up. The  power spectrum  ${\mathcal P}_{h}(\eta, k)$ is defined as
\begin{equation}
\label{eq:Ph}
\left\langle h_{\mathbf{k}}(\eta) h_{\mathbf{p}}(\eta)\right\rangle= \frac{2 \pi^{2}}{k^{3}} \delta^{3}(\mathbf{k}+\mathbf{p}) {\mathcal P}_{h}(\eta, k).
\end{equation}

Using the Green's function method, the solution to the equation of $h_{\mathbf{k}}$ is
\begin{equation}
 h_{\mathbf{k}}(\eta)=\frac{1}{a(\eta)}\int^{\eta} \mathrm{d} \overline{\eta} G_{\mathbf{k}}(\eta, \overline{\eta}) a(\overline{\eta}) S_{\mathbf{k}}(\overline{\eta}),
\end{equation}
where the Green's function $G_{\mathbf{k}}(\eta, \overline{\eta})$ satisfies
\begin{equation}
G_{\mathbf{k}}^{\prime \prime}(\eta, \overline{\eta})+\left(k^{2}-\frac{a^{\prime \prime}(\eta)}{a(\eta)}\right) G_{\mathbf{k}}(\eta, \overline{\eta})=\delta(\eta-\overline{\eta}),
\end{equation}
with the primes denote derivatives with respect to $\eta$. In the RD era, which is
\begin{equation}
\begin{aligned} G_{\mathbf{k}}(\eta, \overline{\eta}) &=\frac{1}{k}\sin (k \eta-k \overline{\eta}).\end{aligned}
\end{equation}

Assuming  $\psi_{\mathbf{k}}$ is Gaussian, the four-point correlation function of $\psi_{\mathbf{k}}$ can be transformed into the two-point correlation function. Then  we obtain the power spectrum by introduce three dimensionless variables $u \equiv|\mathbf{k}-\mathbf{p}| / k$, $v \equiv|\mathbf{p}| / k$ and  $x \equiv k \eta$,~\cite{ref13}
\begin{equation}
\label{eq:twopoint2}
{\mathcal P}_{h}(\eta, k)=4 \int_{0}^{\infty} d v \int_{|1-v|}^{1+v} d u\left(\frac{4 v^{2}-\left(1+v^{2}-u^{2}\right)^{2}}{4 u v}\right)^{2} \mathcal{I}^{2}(x, u, v) \PP (k u) \PP (k v).
\end{equation}
Consider  the late-time limit $x \rightarrow \infty$, the term $\mathcal{I}(x, u, v)$ in the RD era can be calculated as
\begin{equation}
\begin{aligned} \mathcal{I}_{RD}(x \rightarrow \infty, u, v)=\frac{3\left(u^{2}+v^{2}-3\right)}{4 u^{3} v^{3} x}\Big\{&\sin x\Big[-4 u v+\left(u^{2}+v^{2}-3\right) \log \left|\frac{3-(u+v)^{2}}{3-(u-v)^{2}}\right|\Big] \\ &-\pi\left(u^{2}+v^{2}-3\right) \Theta(v+u-\sqrt{3}) \cos x\Big\}.\end{aligned}
\end{equation}
After take the oscillation average, one can obtain
\be
&& \overline{\mathcal{I}_{RD}^{2}(x \rightarrow \infty, u, v)}
 = \dfrac{1}{2}\left(\frac{3}{4 u^{3} v^{3} x}\right)^{2}\left(u^{2}+v^{2}-3\right)^{2} \nonumber \\
&& \quad \Big\{ \Big[-4 u v+\left(u^{2}+v^{2}-3\right) \ln \left|\frac{3-(u+v)^{2}}{3-(u-v)^{2}}\right|\Big]^{2}
 +\left[\pi\left(u^{2}+v^{2}-3\right) \Theta(u+v-\sqrt{3})\right]^{2} \Big\}.
\label{eq:I3}
\ee
 Together with Eq.~(8) and Eq.~(13), and using $\mathcal{H}=1/\eta$ in the radiation-dominated era, we finally get the energy spectrum of GWs
\be
&& \Omega_{\mathrm{GW}}(\eta,k)
 = \dfrac{1}{12}\int_{0}^{\infty} dv \int_{|1-v|}^{1+v} du \left(\frac{4 v^{2}-\left(1+v^{2}-u^{2}\right)^{2}}{4 u v}\right)^{2} \PP (k u) \PP (k v) \nonumber \\
&& \quad \left(\frac{3}{4 u^{3} v^{3} }\right)^{2}\left(u^{2}+v^{2}-3\right)^{2} \nonumber \\
&& \quad \Big\{\Big[-4 u v+\left(u^{2}+v^{2}-3\right) \ln \left|\frac{3-(u+v)^{2}}{3-(u-v)^{2}}\right|\Big]^{2}+\left[\pi\left(u^{2}+v^{2}-3\right) \Theta(u+v-\sqrt{3})\right]^{2} \Big\}.
\label{eq:final}
\ee
\section{The models}


\subsection{Model I: A polynomial potential with inflection points\label{sec:infl}}
Motivated by an effective field theory with a cutoff scale $\Lambda$, a polynomial potential can be generally given by\cite{Bhaumik:2019tvl,Enqvist:2003gh,Burgess:2009ea,Marchesano:2014mla,ref37,ref38}
\begin{equation}
V_{\mathrm{eff}}(\phi)=\sum_{n=0} \frac{b_{n}}{n !}\left(\frac{\phi}{\Lambda}\right)^{n}.
\end{equation}
We ignore the constant term and first order term to make the potential and it's first-order derivative vanish at the origin.
In order to build a model with three inflection points, where both the first and second order derivatives of $V$ vanish, there must have six free parameters, 
thus we truncate the effective potential to the eight-order.
Then the potential can be parameterized as
\begin{eqnarray}
V_{\mathrm{eff}}(\phi)=V_{0}\left[\frac{c_{2}}{2 !}\left(\frac{\phi}{\Lambda}\right)^{2}+\frac{c_{3}}{3 !}\left(\frac{\phi}{\Lambda}\right)^{3}+\frac{c_{4}}{4 !}\left(\frac{\phi}{\Lambda}\right)^{4}+\frac{c_{5}}{5 !}\left(\frac{\phi}{\Lambda}\right)^{5}+\frac{c_{6}}{6 !}\left(\frac{\phi}{\Lambda}\right)^{6}+\frac{c_{7}}{7 !}\left(\frac{\phi}{\Lambda}\right)^{7}+\frac{1}{8 !}\left(\frac{\phi}{\Lambda}\right)^{8}\right], \nonumber\\
\end{eqnarray}
where $V_0$ is an overall factor, which can be constrained by the amplitude of scalar perturbations $A_s$, and $c_{2-7}$ are six free parameters. 

Although in some choices of parameter space, such potential allows the existence of inflection points, however it is non-renormalizable.
So here we introduce an appropriate factor and then the potential becomes
\begin{equation}
V(\phi)=\frac{V_{\mathrm{eff}}(\phi)}{\left(1+\xi \phi^{2}\right)^{2}}.
\end{equation}
Such factor is usually arose from the scalar field non-minimal coupling to gravity\cite{Bhaumik:2019tvl,ref38}.
In some choices of parameter space, the potential $V(\phi)$  can generate three inflection points which give an approximate scale invariant spectrum at CMB scales, and at the same time generate two large peaks in the power spectrum at small scales to induce double peaks GW spectrum.

For convenience of discussion, we assume that the potential have three inflection point at $\phi_i (i=1,2,3)$ respectively, where $V'(\phi_i)=0$ and  $V''(\phi_i)=0$. More generally, in order to study the slight deviations from a perfect inflection point, we introduce three more free parameters $\alpha_i$ and set $V'(\phi_i)=\alpha_i$ and  $V''(\phi_i)=0$.  Then the parameters $c_{2\sim7}$ in Eq.(18) can be expressed as functions of $\phi_i$ and $\alpha_i$. By fine-tuning the parameters, one can get a potential with three inflection points, which is consistent with the CMB constraints and induce a second-order GW spectrum with double peaks.

For instance, we take the following parameter set
\begin{eqnarray}
&&V_0=6.96\times10^{-11}, \;\;\; \Lambda=0.5, \;\;\;\xi= 0.4, \nonumber\\
&&\phi_1=0.49,\ \ \ \ \ \alpha_1= -8.485824\times10^{-6},\nonumber\\
&& \phi_2=1.09, \ \ \ \ \ \ \alpha_2= -2.104572\times10^{-5},\nonumber\\
&& \phi_3=2, \ \ \ \ \ \ \ \ \ \alpha_3=7.2\times10^{-7},
\end{eqnarray}
and the corresponding potential are show in Fig.1.
\begin{figure}
 \begin{minipage}[t]{0.49\linewidth} 
	\centering
	\includegraphics[width=.99\textwidth]{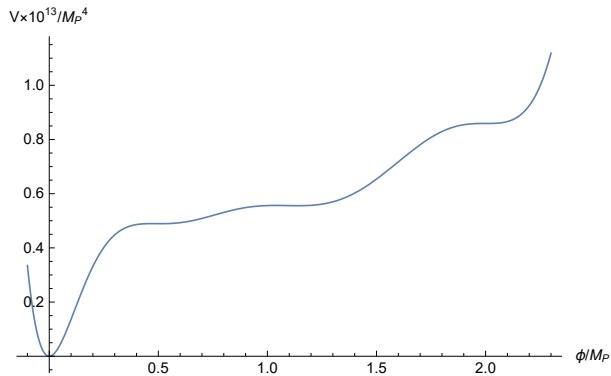}
	\label{fig:a}
 \end{minipage}
 \caption{Scalar potential $V(\phi)$ of Model I for the parameter set (20).}
\label{fig:1}
\end{figure}
We can see that there are three inflection points. The inflation starts near the first inflection point at high scales and leads to a nearly scale-invariant spectrum, then slowly rolls down the smooth plateau-like regions with a nearly constant Hubble friction. Whenever the inflaton meets a cliff, it speeds up quickly, until it reaches the next inflection point plateau, where the Hubble friction rapidly slows it down again. Such point leads to a phase of ultra-slow-roll last about $20$ e-folding numbers, which generates a large peak in the power spectrum. Similarly, near the last inflection point, the inflaton becomes ultra-slow-roll again and generates the second peak of the power spectrum.
\subsection{Model II: A Higgs like potential with the running of quartic coupling \label{sec:infl}}


In this subsection, we shall consider a Higgs-like potential. During inflation the field values is large, so only the quartic part of the potential matters. Although the $\phi^4$ model is ruled out by the CMB observations, however, if one consider the interactions between the inflaton and other fields, required
for reheating, the situation will be changed\cite{Ballesteros:2015noa}. The radiative corrections can be accounted for through the running of the quartic coupling, then the potential is approximated as
\begin{eqnarray}
&&V(\phi)=\frac{\lambda(\phi)}{4!}\phi^4,
\end{eqnarray}
with $\lambda(\phi)$ is an effective field-dependent coupling, it can be parameterized as\cite{Ballesteros:2015noa,ref21}
\begin{eqnarray}
&&\lambda=\lambda_0\Big[1+b_1\ln\Big(\frac{\phi^2}{\phi_0^2}\Big)+b_2\ln\Big(\frac{\phi^2}{\phi_0^2}\Big)^2+b_3\ln\Big(\frac{\phi^2}{\phi_0^2}\Big)^3
+b_4\ln\Big(\frac{\phi^2}{\phi_0^2}\Big)^4+b_5\ln\Big(\frac{\phi^2}{\phi_0^2}\Big)^5+b_6\ln\Big(\frac{\phi^2}{\phi_0^2}\Big)^6\Big], \nonumber\\
\end{eqnarray}
where the overall factor $\lambda_0$ can be constrained by the amplitude of scalar perturbations $A_s$, and the logarithms comes from two-loop and higher order terms in the Coleman-Weinberg
expansion\cite{ref39}. In order to make the potential have three inflection points, we truncate it to six-order here.

Similarly as in Model I, we assume that the potential have three inflection points at $\phi_1, \phi_2$ and $\phi_3$, respectively, and set $V'(\phi_i)=\beta_i$ and  $V''(\phi_i)=0$  to study the slight deviations from a perfect inflection point.  Then the six parameters $b_{1\sim6}$ in Eq.(22) can be expressed as functions of $\phi_i$ and $\beta_i$. For some parameter sets, the potential $V(\phi)$  can also generate three inflection points which have the predictions in good agreement with the current CMB measurements and induce GWs energy spectrum with  double peaks. 
For instance, we take the following parameter set
\begin{eqnarray}
&&\lambda_0=0.8983\times10^{-13}, \ \ \ \phi_0= 1.7, \nonumber\\
&&\phi_1=0.473,\ \ \ \ \ \ \  \beta_1=-0.0501578,\nonumber\\
&& \phi_2=1.1,\ \ \ \ \ \ \ \  \beta_2= -0.2150073,\nonumber\\
&& \phi_3=2,\ \ \ \ \ \ \ \ \ \ \beta_3=0.00355,
\end{eqnarray}
and the corresponding potential are show in Fig.2.

\begin{figure}
 \begin{minipage}[t]{0.49\linewidth} 
	\centering
	\includegraphics[width=.99\textwidth]{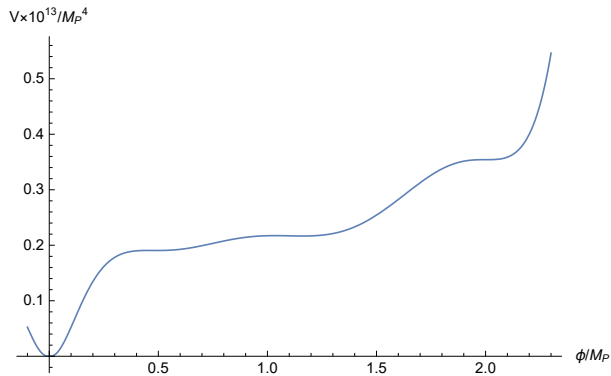}
	\label{fig:a} %
 \end{minipage}

 \caption{Scalar potential $V(\phi)$ of Model II for the parameter set (23).}
\label{fig:1}
\end{figure}

\section{Numerical Results}
In this section, we shall numerically calculate the power spectrum of scalar perturbations and the energy spectrum of induced GWs in Model I for parameter set (20) and in Model II for parameter set(23), respectively, then compare the results of $\Omega_{\mathrm{GW},0}$ to the expected sensitivity curves of several planned GW detectors. We also calculate the abundance of PBHs produced in the two models
using the Press-Schechter approach\cite{ref400} of gravitational collapse.
\subsection{Inflaton dynamics}
It has been pointed out in several references that near the inflection point the potential becomes extremely flat, thus the slow-roll approximation fails and is superseded by the ultra-slow-roll trajectory ~\cite{ref40,ref41,ref42,ref43}. So one must use the Hubble slow-roll parameters instead the traditional one~\cite{ref44,ref45,ref46}, which is defined as
\begin{eqnarray}
&&\epsilon_H=-\frac{\dot{H}}{H^2},\nonumber\\
&&\eta_H=-\frac{\ddot{H}}{2H\dot{H}}=\epsilon_H-\frac{1}{2}\frac{d\ln\epsilon_H}{dN_e},
\end{eqnarray}
with dots represent derivatives with respect to cosmic time, and $N_e$ denotes the $e$-folding numbers.
In Fig.3, we plot the Hubble slow-roll parameters $\epsilon_H$ and $\eta_H$ as functions of $N_e$ for the two models, respectively.
\begin{figure}
 \begin{minipage}[t]{0.49\linewidth} 
	\centering
	\includegraphics[width=.99\textwidth]{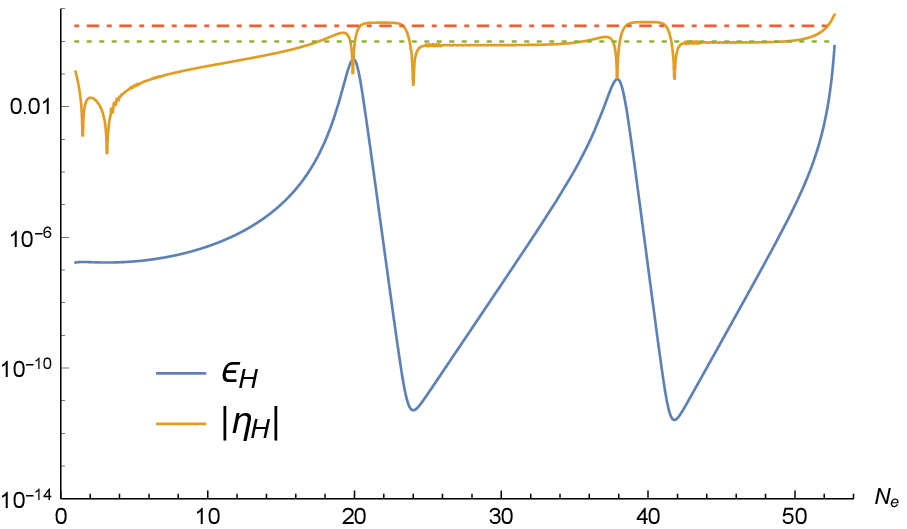}
	\label{fig:a} %
 \end{minipage}
 \begin{minipage}[t]{0.49\linewidth} 
	\centering
	\includegraphics[width=.99\textwidth]{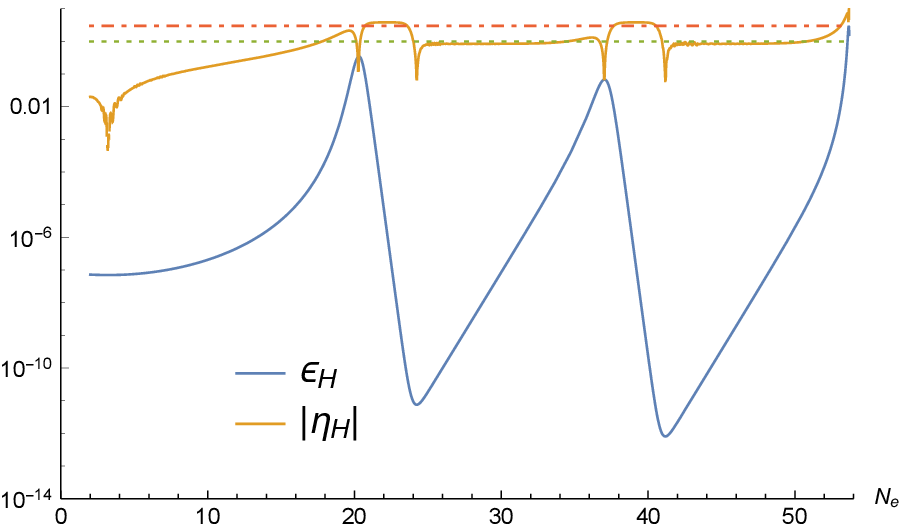}
	\label{fig:b}
 \end{minipage}
 \caption{Hubble slow-roll parameters $\epsilon_H$ and $\eta_H$  as functions of the e-folding number $N_e$ in Model I for parameter set (20) (left panel), Model II for parameter set (23) (right panel). The dashed lines indicate the values $1$ and $3$ }
\label{fig:1}
\end{figure}
We can see that near the inflection points the Hubble slow-roll parameter $|\eta_H|>3$, so the inflation becomes ultra-slow-roll, which lead to large valleys on the curve of $\epsilon_H$ last about $15-20$ e-folding numbers, and will give rise to double peaks in the scalar power spectrum.

The scalar spectral index as well as the tensor-to-scalar ratio can be expressed using $\epsilon_H$ and $\eta_H$  as
\begin{eqnarray}
&&n_s=1-4\epsilon_H+2\eta_H,\nonumber\\
&&r=16\epsilon_H.
\end{eqnarray}
For the two models, the numerical results of $n_s$ and $r$ are present in Tab.1. The amplitude of the primordial curvature perturbations $A_s$ and the e-folding numbers during inflation $N_e$ are also list there.
\begin{table}

\begin{tabular}{c||c|c|c|c|c|c|c}
&$n_s$&$r$&$\ln(10^{10}A_s)$&$ N_e$\\
\hline  
Model I&$0.968655$&$2.765\times10^{-6}$&$3.0442$&$53.2$\\

Model II&$0.966655$&$1.140\times10^{-6}$&$3.0439$&$53.7$\\


\end{tabular}
\caption{The numerical results for the two models.}
\end{table}
The results are in agreement with the  current  CMB constraints $n_s=0.9649\pm0.0042$, $r<0.064$ and  $\ln(10^{10}A_s)=3.044\pm0.014$ from Planck 2018 \cite{ref6}.

Since the slow-roll approximation fails near the inflection point, the calculation of the scalar perturbations using the approximate expression $\PP \simeq\frac{1}{8\pi^2M_P^2}\frac{H^2}{\epsilon_H}$
will underestimates the power spectrum~\cite{ref21,ref25}. Thus it is necessary to solve the Mukhanov-Sasaki(MS) equation of mode function numerically,
and then the power spectrum can be calculated by
\begin{eqnarray}
&&\PP=\frac{k^3}{2\pi^2}\Big|\frac{u_k}{z}\Big|^2_{k\ll\mathcal{H}}.
\end{eqnarray}
The numerical results of scalar power spectrum for the two models are shown in Fig.4. The blue line is the numerical result of MS equation and the orange line is the approximate result.
\begin{figure}
 \begin{minipage}[t]{0.49\linewidth} 
	\centering
	\includegraphics[width=.99\textwidth]{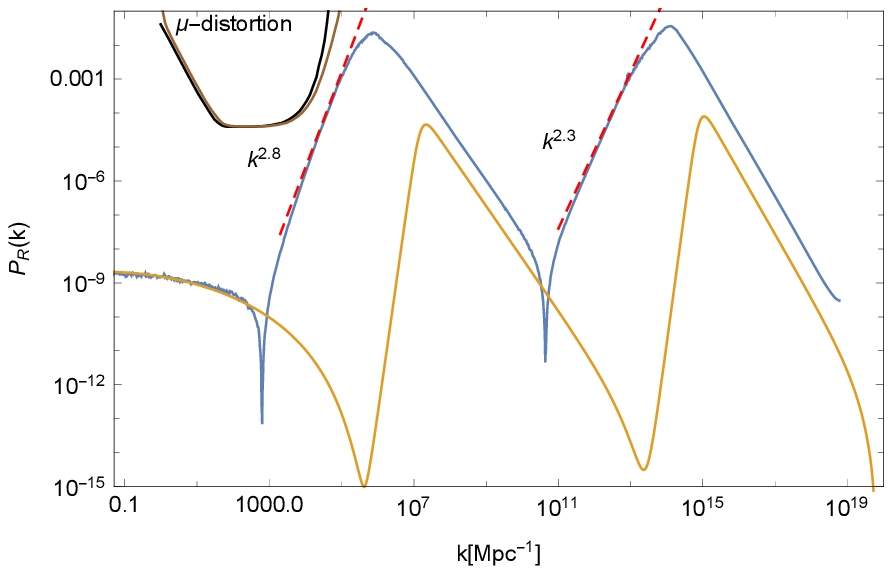}
	\label{fig:a} %
 \end{minipage}
  \begin{minipage}[t]{0.49\linewidth} 
	\centering
	\includegraphics[width=.99\textwidth]{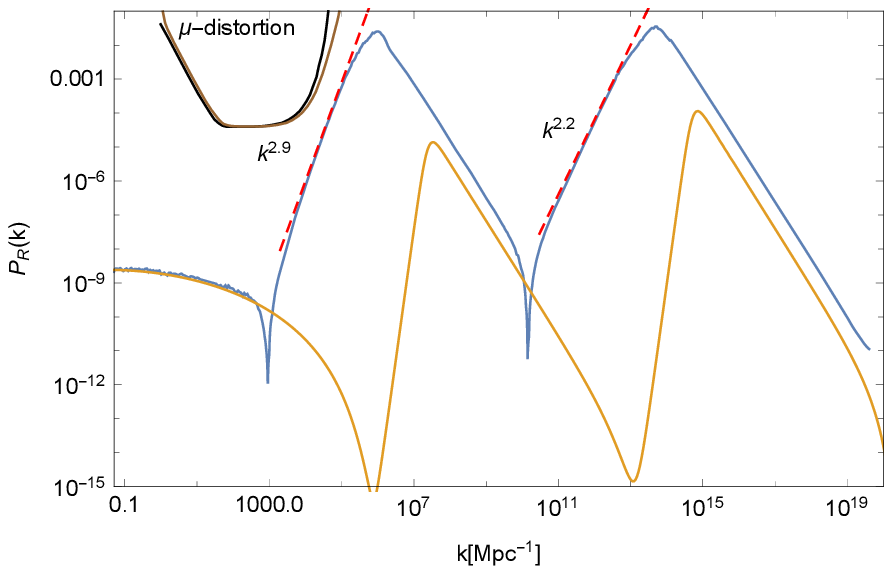}
	\label{fig:b}
 \end{minipage}

 \caption{Primordial power spectrum of scalar perturbations predicted by Model I for parameter set (20) (left panel), and by Model II for parameter set (23) (right panel). And the upper bound from $\mu$-distortion for a delta function power spectrum(Black line) and for the steepest growth $k^4$ power spectrum(Brown line)\cite{Byrnes:2018txb}}
\label{fig:1}
\end{figure}
We can see that for both models, the scalar power spectrum matches the Planck observations at CMB scales, and has two large peaks at small scales with a height of about seven orders of magnitude more than the spectrum at CMB scales, such peaks will lead to the production of non-negligible second-order GWs with double peaks energy spectrum.

It is interesting  to note that in Ref.\cite{Byrnes:2018txb}, the authors pointed out that in canonical single field inflation, the steepest growth index of the power spectrum before the peak is $k^4$, and we find that in our models the growth index is $~k^{2.8}$ and $~k^{2.3}$ in Model I and $k^{2.9},k^{2.2}$ for the peaks in Model II (the red dashed line of Fig.4), which are all slower than $k^4$. And the constraints to the primordial power spectrum  from $\mu$-distortion are also show in Fig.4.

\subsection{Energy spectrum of induced GWs}
Using the scalar power spectrum obtained in the previous subsection, and consider that the GW energy spectrum at the present time $\Omega_{\mathrm{GW}, 0}$ is related to the one produced in the RD era as\cite{ref33}
\begin{equation}
\label{eq:Omegaappr}
\Omega_{\mathrm{GW}, 0}=0.83 \left(\frac{g_{*,0}}{g_{*,p}}\right)^{-1 / 3} \Omega_{r, 0} \Omega_{\mathrm{GW}},
\end{equation}
with $\Omega_{r, 0}\simeq9.1\times10^{-5}$ is the current density fraction of radiation,
$g_{*,0}$ and $g_{*,p}$ are the effective degrees of freedom for energy density at the present time
and at the time when the peak mode crosses the horizon, respectively.
We numerically calculate the energy spectrum of induced GWs for the two models and show them as functions of the present value of the frequency $f$ in Fig.5, with
\begin{equation}
\label{eq:Omegaappr}
f \approx 0.03 \mathrm{Hz} \frac{k}{2\times 10^7 \mathrm{pc}^{-1}}.
\end{equation}
And the sensitivity curves of several planned
GW detectors are also shown there~\cite{ref52,ref53,ref54,ref55,ref56,ref24}.
\begin{figure}
\begin{minipage}[t]{0.49\linewidth} 
	\centering
	\includegraphics[width=.99\textwidth]{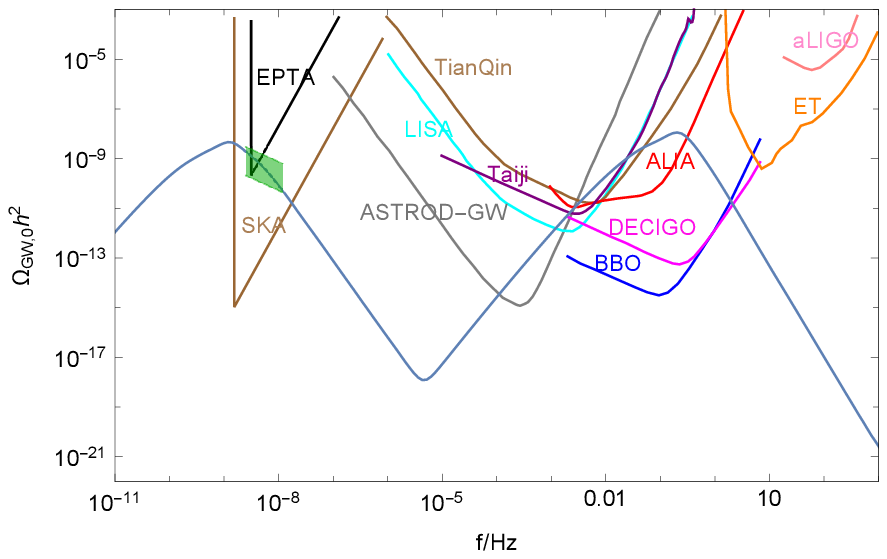}
	\label{fig:a} %
 \end{minipage}
  \begin{minipage}[t]{0.49\linewidth} 
	\centering
	\includegraphics[width=.99\textwidth]{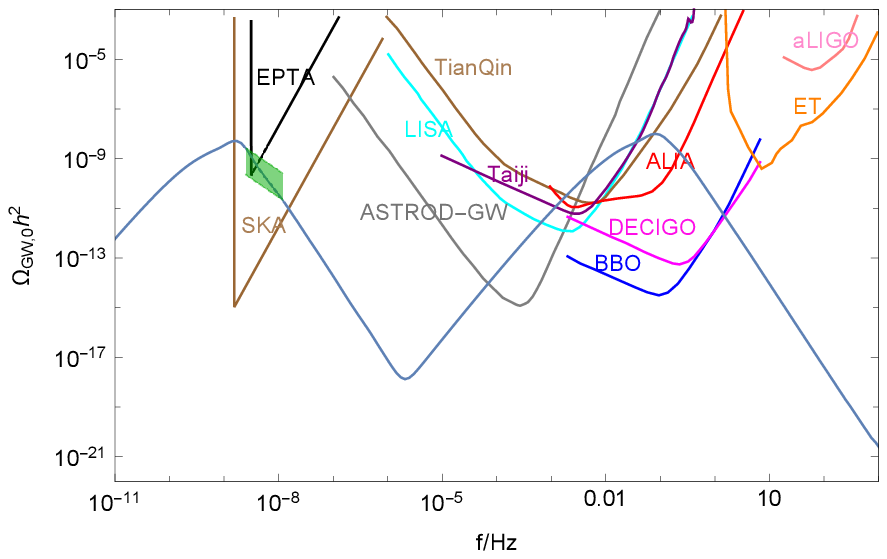}
	\label{fig:b}
 \end{minipage}
\caption{Energy spectrum of the induced GWs at the present time
predicted by Model I for parameter set (20) (left panel), and by Model II for parameter set (23) (right panel).
The curves in the upper part are the expected sensitivity curves of the Square Kilometer Array (SKA), European Pulsar Timing Array (EPTA), Astrodynamical Space
Test of Relativity using Optical-GW detector (ASTROD-GW),Taiji, Laser Interferometer Space Antenna (LISA), TianQin, Advanced Laser Interferometer Antenna (ALIA), Big Bang Observer (BBO), Deci-hertz Interferometer GW Observatory (DECIGO),Einstein Telescope (ET),  Advanced LIGO (aLIGO), respectively. These sensitivity curves are taken from Ref.~\cite{ref52,ref53,ref54,ref55,ref56,ref24} The green region  show the  $2 \sigma$ confidence level of the NANOGrav results with the tilt of $5-\gamma=-1$(left panel) and $5-\gamma=-1.5$(right panel)\cite{Arzoumanian:2020vkk}.
}
\label{fig:gwLisa}
\end{figure}

We can see that the energy spectrum of induced GWs have double peaks. The peak at low frequency, about $10^{-9}-10^{-8}Hz$, lies above the expected sensitivity curves of SKA and EPTA for both models, and the peak at high frequency, about $0.01-1Hz$, is within the frequency range of LISA~\cite{ref52}, ALIA, Taiji~\cite{ref53} and TianQin~\cite{ref55}, and the energy spectrum curves lies above the expected sensitivity curves of them. So such kind of GWs can be detected in near future.

Recently, the NANOGrav collaboration has published an analysis of 12.5-year of PTA, which can be explained by the stochastic GWs\cite{Arzoumanian:2020vkk,Vaskonen:2020lbd,Kohri:2020qqd,DeLuca:2020agl,Kawasaki:2021ycf}. The potential GW signal can
be fitted by a power-law spectrum around $f_{yr}\simeq3.1\times10^{-8}$Hz,
\begin{equation}
\Omega_{GW}(f)=\frac{2\pi^2f_{yr}^2}{3H_0^2}A^2_{GWB}\Big(\frac{f}{f_{yr}}\Big)^{5-\gamma},
\end{equation}
where $H_0\equiv100 h$ km/s/Mpc and the exponent $5-\gamma\in(-1.5,0.5)$ at $1\sigma$ confidence level. In Fig.5 we also show the observed GWs for $5-\gamma=-1$(left panel) and $5-\gamma=-1.5$(right panel) with $2 \sigma$ uncertainty on $A_{GWB}$. We can see that the peak of induced GW spectrum at frequencies around nanohertz can also explain the NANOGrav signals.

In addition, since the energy spectrum have two peaks, so if one of the peaks at nHz is detected by observers, the other  peak at $0.01$Hz should be detected by other detectors like LISA in the future. Thus it is different from the models with only one peak, which can only be detected in one frequency ranges. So the double peaks models can be distinguished with other single peak models.

\subsection{Production of primordial black holes \label{sec3}}
When the large amplitude of primordial fluctuations re-enters the Hubble horizon after inflation, it will undergoes gravitational collapse and form PBHs if the fluctuation is significantly large.
The mass of the resulting PBHs is assumed to be proportional to the horizon mass at re-entry time,
\begin{eqnarray}
&&M=\gamma M_H=\gamma\frac{4}{3}\pi \rho H^{-3},
\end{eqnarray}
which  can be approximated as\cite{ref500}
\begin{eqnarray}
&&M\simeq10^{18} \text{g} \Big(\frac{\gamma}{0.2}\Big)\Big(\frac{g_*}{106.75}\Big)^{-1/6}\Big(\frac{k}{7\times10^{13}\text{Mpc}^{-1}}\Big)^{-2},
\end{eqnarray}
where $\gamma\sim 0.2$ depends on the details of the gravitational collapse\cite{ref501}, and $g_*\sim 106.75$ is the effective degrees of freedom for energy density.

In the context of the Press-Schechter model of gravitational collapse, assuming that the probability distribution of density perturbations are Gaussian with width $\sigma(M)$, the mass fraction in PBHs of mass $M$ is given by
\begin{eqnarray}
\beta(M)\equiv\frac{\rho_{PBH}(M)}{\rho_{tot}}&&=\frac{1}{\sqrt{2 \pi \sigma^2(M)}}\int^{\infty}_{\delta_c}d\delta \text{exp}\Big(\frac{-\delta^2}{2 \sigma^2(M)}\Big)\nonumber\\
&&=\frac{1}{2}\text{erfc}\Big(\frac{\delta_c}{\sqrt{2\sigma^2(M)}}\Big),
\end{eqnarray}
where $\delta_c\simeq0.45$ is the threshold for collapse\cite{ref502,ref503}. And the variance of the comoving density perturbations $\sigma^2(M)$ is assumed  to be coarse-grained at a scale $R=1/k$, during the radiation-dominated era, which is given by,
\begin{eqnarray}
&&\sigma^2(M(k))=\frac{16}{81}\int\frac{dq}{q}(qR)^4P_{R}(q)W(qR)^2,
\end{eqnarray}
with the smoothing window function $W(x)$ is taken to be a Gaussian $W(x)=exp(-x^2/2)$.
Then integrating over all masses $M$ one can get the  present abundance of PBHs,
\begin{eqnarray}
&&\Omega_{PBH}=\int\frac{dM}{M}\Omega_{PBH}(M),
\end{eqnarray}
with
\begin{eqnarray}
&&\frac{\Omega_{PBH}(M)}{\Omega_{DM}}=\Big(\frac{\beta(M)}{1.6\times10^{-16}}\Big)\Big(\frac{\gamma}{0.2}\Big)^{3/2}\Big(\frac{g_*}{106.75}\Big)^{-1/4}\Big(\frac{M}{10^{18} \text{g} }\Big)^{-1/2},
\end{eqnarray}
where $\Omega_{DM}\simeq0.26$ is the total dark matter abundance\cite{ref504}.


The numerical results show that the two peaks models can give rise to two PBHs populations with different masses. The results of the PBHs mass and abundance are present in Tab.2.
\begin{table}

\begin{tabular}{c||c|c|c|c}
&$ M_{PBHs}^{peak}/M_{\odot}$&$ \Omega_{PBH}/\Omega_{DM}$&$ M_{PBHs}^{peak}/M_{\odot}$&$ \Omega_{PBH}/\Omega_{DM}$\\
\hline  
Model I&$7.490$&$1.568\times10^{-17}$&$2.818\times10^{-16}$&$0.768$\\

Model II&$4.883$&$4.561\times10^{-17}$&$1.735\times10^{-15}$&$0.031$\\


\end{tabular}
\caption{The numerical results of PBHs production for the two models.}
\end{table}
We can see that only one type of the PBHs with mass around $10^{-16}\sim10^{-15}M_{\odot}$ can be a significant fraction of dark matter, and the abundance of the others mass is two small. The abundance of PBHs for the two models and the observational constraints from Ref.\cite{ref37} are show in Fig.6




\begin{figure}\small
\begin{minipage}[t]{0.49\linewidth} 
	\centering
	\includegraphics[width=.99\textwidth]{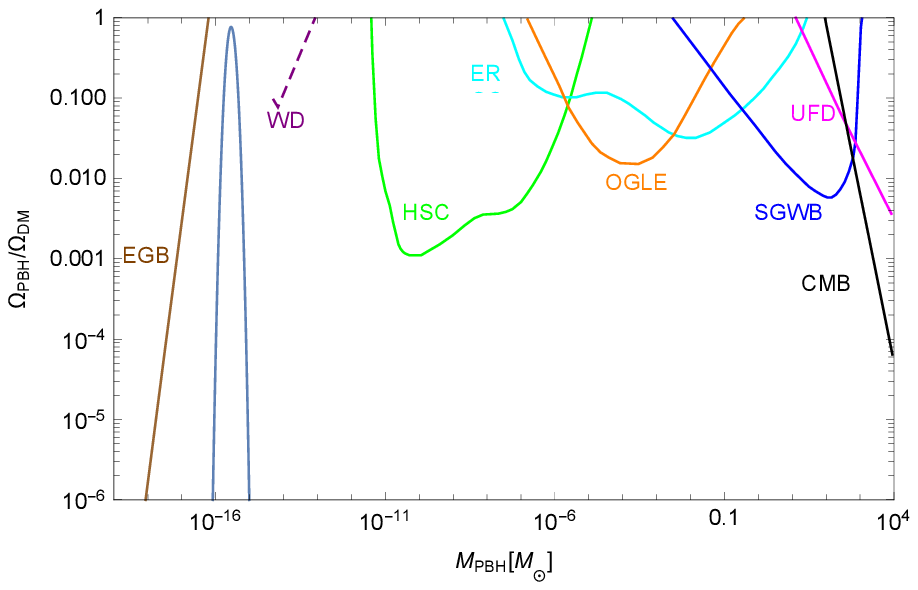}
	\label{fig:a} %
 \end{minipage}
  \begin{minipage}[t]{0.49\linewidth} 
	\centering
	\includegraphics[width=.99\textwidth]{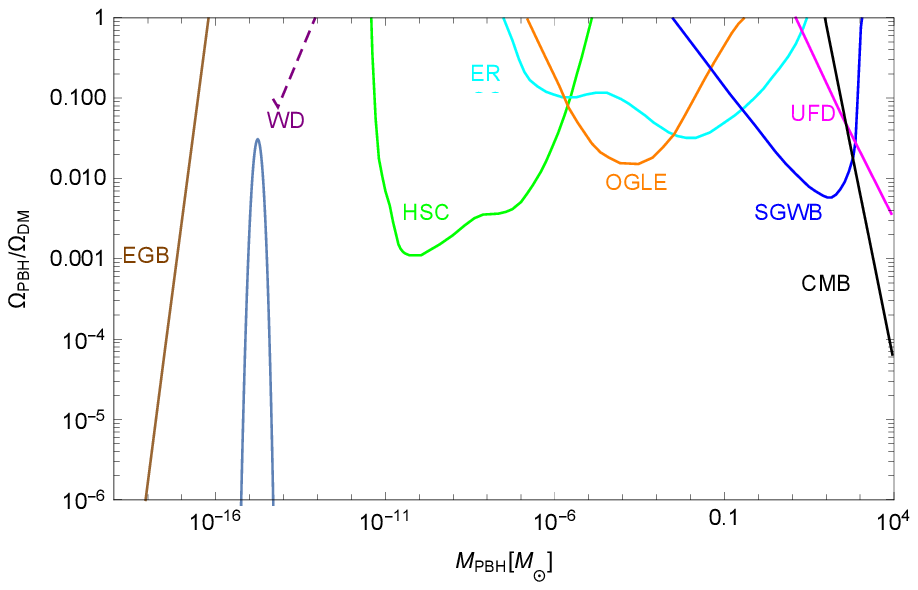}
	\label{fig:b}
 \end{minipage}

     \caption{The abundance of PBHs produced in Model I for parameter set (20) (left panel), and in Model II for parameter set (23) (right panel). The curves in the upper part are the observational constraints on the PBH abundance, which are taken from Ref.\cite{ref37}.}
    \label{fig1}
\end{figure}

\section{Summary \label{sec:infl}}

In this paper, we investigated the possibility to induce double peaks of GW spectrum from single-field inflationary models with inflection points. We found that such double peaks spectrum can be realized by the polynomial potential from effective field theory with a cut off scale(Modle I) or realized by the Higgs like $\phi^4$ potential with the running of quartic coupling from radiative corrections(Model II). In order to generate an inflationary potential with three inflection points, and make the predictions to fall in the observational windows, the fine-tuning of several parameters and the initial condition on the scalar field is necessary. We find that for some choices of parameter sets, the inflection point at large scales make the prediction of scalar spectral index and tensor-to-scalar ratio consistent with the current CMB constraints, and the other two inflection points generate two large peaks in the power spectrum at small scales to induce GWs with double peaks spectrum. We calculated the energy spectrum of GWs numerically and shown that the peak at low frequency $10^{-9}-10^{-8}Hz$ lies above the expected sensitivity curves of SKA and EPTA, which can explain the NANOGrav signal. The peak at high frequency $0.01-1Hz$ lies above the expected sensitivity curves of LISA, ALIA, Taiji, TianQin, etc, so it can be detected in near further. In addition, since the double peaks energy spectrum can be detected by observers in different frequency ranges thus it can be distinguish with other single peak models.
Moreover, we also calculated the abundance of PBHs produced in the two models using the Press-Schechter approach\cite{ref400} of gravitational collapse, and found that the PBHs with the mass around $10^{-16}\sim10^{-15}M_{\odot}$ can account for a significant fraction of dark matter.

\begin{acknowledgments}
This work was supported by ``the National Natural Science Foundation of China'' (NNSFC) with Grant No. 11705133. XYY was supported by ``the Department of education of Liaoning province '' with Grant No.2020LNQN14.
\end{acknowledgments}

\end{document}